\documentclass[11pt, a4paper]{article}

\usepackage[utf8]{inputenc}
\usepackage[english]{babel}
\usepackage[margin=2.3cm]{geometry} 
\usepackage{graphicx}
\usepackage{amsmath,amsfonts,amssymb,amsthm}
\usepackage{hyperref}
\usepackage{xspace}
\usepackage{placeins} 
\usepackage{tcolorbox} 
\usepackage[numbers]{natbib} 
\usepackage{lineno}	
\usepackage{authblk} 
\usepackage{appendix} 
\usepackage{color}
\usepackage{lineno}
\usepackage{mathrsfs}
\usepackage{bm}
\usepackage{dsfont}
\usepackage{wasysym}
\usepackage{textgreek}
\usepackage{multirow}
\usepackage{rotating}
\usepackage{booktabs,caption}
\usepackage{threeparttable}
 \usepackage{setspace} 
  
\newcommand{\papertitle}{A thermo-hygro computational model to determine the factors dictating cold joint formation in 3D printed concrete}

\def\degC{$^{\circ}$C~}

\newcommand{\reffig}[1]{Figure~\ref{fig:#1}}

\newcommand{\refeq}[1]{Eq.~(\ref{eq:#1})}

\newcommand{\reftab}[1]{Table~\ref{tab:#1}}

\newcommand{\refsec}[1]{Section~\ref{sec:#1}}
\newcommand{\refsecs}[2]{Sections~\ref{sec:#1}, \ref{sec:#2}}
\begin{document}

\title{\papertitle}
\author{Michal~Hlobil}
\author{Luca~Michel}
\author{Mohit~Pundir}

\author{David S.\ Kammer}
\affil{Institute for Building Materials, ETH Zurich, Switzerland}

\maketitle

\section*{Abstract}
Cold joints in extruded concrete structures form once the exposed surface of a deposited filament dries prematurely and gets sequentially covered by a layer of fresh concrete. This creates a material heterogeneity which lowers the structural durability and shortens the designed service life. Many factors concurrently affect cold joint formation, yet a suitable tool for their categorization is missing. Here, we present a computational model that simulates the drying kinetics at the exposed structural surface, accounting for cement hydration and the resulting microstructural development. The model provides a time estimate for cold joint formation as a result. It allows us to assess the drying severity for a given structure's geometry, its interaction with the environment, and ambient conditions. We evaluate the assessed factors and provide generalized recommendations for cold joint mitigation.

\section*{Keywords}
3D printed concrete; Additive manufacturing; Cold joint; Hydration modeling; Multiphysics modeling

\newpage
\section{Introduction}
\begin{figure}[htb!]
    \centering
    \includegraphics[scale=0.9]{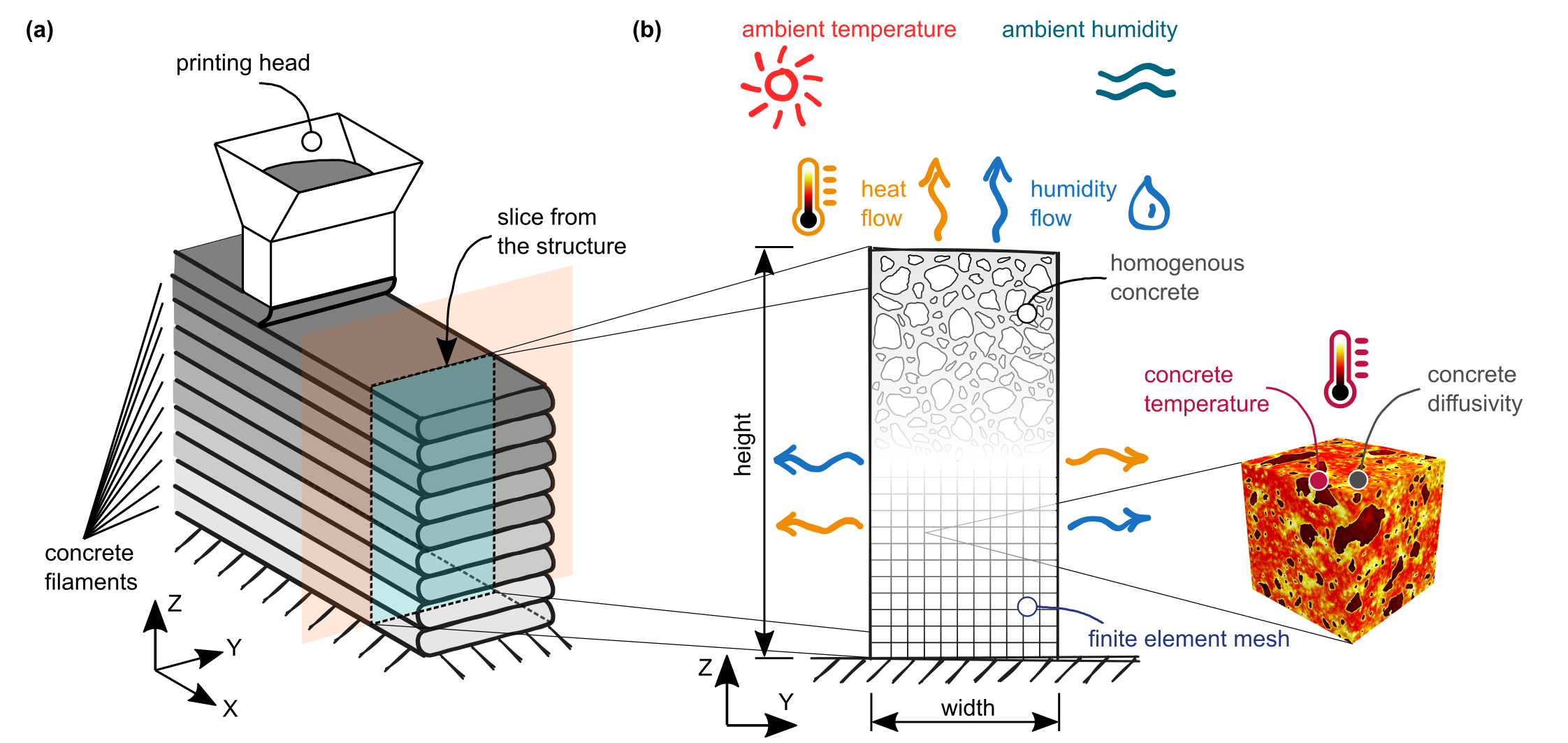}
	\caption{Schematic illustration of the modeling approach. (a) Sketch of a wall-like extruded concrete structure consisting of sequentially stacked filaments. (b) A planar 2D slice from the structure comprising a materially homogeneous cross-section. A finite element mesh is created over the entire domain to simulate transient and coupled heat and humidity transport, leading to a prediction of surface drying rate. The model accounts for i) fresh concrete hardening, accompanied by internal heat generation, which leads to a temperature increase within the concrete structure, and ii) material diffusivity, capturing the state of the pore network formed within the material.}
	\label{fig:mmodel}
\end{figure}

The adoption of extrusion-based additive manufacturing opened new possibilities for producing concrete structures.
This technology draws from the synergy between computer-aided design, robotics, and material usage optimization~\cite{Lloret-Fritschi:2020, Buswell:2020, Vantyghem:2020}.
Utilizing additive manufacturing, structural elements are produced by successively stacking thin layers, or filaments, of rapidly reacting concrete, which are extruded from a robotically-controlled printing head~\cite{Buswell:2007}, see~\reffig{mmodel}(a).
This technique facilitates the production of shape-optimized elements with complex structural geometries~\cite{Agusti-Juan:2017, Schwartz:2018, Hack:2020}, avoiding traditional formwork.
However, successfully implementing this technology into common practice faces several challenges, both on the structural and material design fronts~\cite{Khan:2020, Bos:2022}.
\FloatBarrier

The material prerequisites for printable concrete call for designing an initially fluid and pumpable mixture that rapidly hardens upon deposition, developing both stiffness and strength.
This ensures the structural stability of the thin-walled extruded structure and sufficient load-bearing capacity to withstand the dead weight of the successively added layers.
However, while rapid strength development of the concrete mix is necessary for 3D printing, it may also negatively affect the bond between two neighboring filaments~\cite{GarciadeSoto:2018, Keita:2019, Wolfs:2019}.

After the extruded filament is deposited, its exposed surface dries, forming a solid crust over time~\cite{Tay:2019}.
Unless the top filament surface is promptly covered with new concrete, the crust thickens and results in a porous region that weakens the interface between the already dry filament and the sequentially added and fully water-saturated filament~\cite{Nerella:2019}.
The presence of this inhomogeneity in the stack of extruded filaments leads to an internal structural defect, commonly denoted as a ``cold joint'' which compromises the long-term durability of the extruded structural element~\cite{Kruger:2021, VanDerPutten:2020, YuZhang:2021, VanDerPutten:2022}.

The origins of crust formation are related to material drying, which inherently arises from two sources.
Water is consumed internally by cement hydration but also evaporates through the unprotected structural surface~\cite{Mechtcherine:2020}, see~\reffig{mmodel}(b).
As the microstructure develops, the smallest pores will be emptied first, leaving voids in the material, which results in self-desiccation.
In addition, cement contact with water triggers a series of spontaneous and highly exothermic reactions, liberating over 500~J as heat per gram of cement over the whole reaction~\cite{Hlobil:2022cementPSD}, as schematically shown in~\reffig{mmodel}(c).
The released thermal energy leads to a progressive temperature increase within the concrete structure~\cite{Azenha:2021}.
Since the cement paste matrix exhibits rather low thermal conductivity, a high-temperature gradient forms between the heating structural core and the exposed outer surface of the concrete filament, accelerating water transport through the pore network.
While all unprotected sides of the extruded structure facilitate water evaporation, the drying of the top surface of the uppermost filament is critical, creating a potential risk for cold joint formation.

Cold joint formation challenges the durability of extruded concrete structures.
Hence, understanding their formation and subsequent mitigation is of great interest.
As shown by Mechtcherine et al.~\cite{Mechtcherine:2020}, multiple input factors influence cold joint formation.
While their experimental quantification is challenging, computational models provide a feasible alternative to studying their magnitude. 
Several multiphysics models have been developed, focusing on predicting the temperature increase in massive concrete structures~\cite{Faria:2006, Smilauer:09, daSilva:2015} and simulating early-age cracking~\cite{Jedrzejewska:2018, Smilauer:2019}.
However, predictive models focusing on drying kinetics of hydrating concrete for extruded structures, which lie at the core of the cold joint formation, are yet missing.

Here, we formulate a computational model to investigate the factors that lead to cold joint formation in extruded concrete structures.
The model is based on solving a coupled thermo-hydro conduction phenomenon over a planar domain, representing a full cross-section of an extruded structure, see~\reffig{mmodel}(b).
The proposed approach captures the kinetics of cold joint formation, accounting for concrete composition, structural dimensions, and interaction with the ambient environment.
We analyze the effect of different factors quantitatively, allowing us to determine the most important ones for cold joint formation, which aids with the material design in practice.

\section{Model formulation}
Origins of cold joint formation point to premature surface drying of hydrating concrete.
This physical phenomenon can be mathematically described by a set of differential equations, formulated here as a coupled thermo-hydro transport problem (the theoretical derivation can be found in Appendix~\ref{sec:DerivationDiffEq}). The governing equations take the form of
\begin{subequations}\label{eq:governing_eq_strong-form}
\begin{align}
    \label{eq:first}
    \nabla \cdot (\lambda_{mat}(\alpha) \, \nabla T_{mat}) + \frac{\partial Q_h}{\partial t} &= \rho_{mat} \, c_{mat}(\alpha) \, \frac{\partial T_{mat}}{\partial t}, \\
    \label{eq:second}
    \nabla \cdot (D_H \, \nabla H_{mat}) - \frac{\partial H_s}{\partial t} + \kappa \, \frac{\partial T_{mat}}{\partial t}&= \frac{\partial H_{mat}}{\partial t},
\end{align}
\end{subequations}
where $\alpha$ denotes the degree of cement hydration, described in~\refsec{Heat_development_during_cement_hydration}, and $T_{mat}$, and $H_{mat}$ the concrete temperature and relative humidity, respectively.
All three unknown quantities then act as field variables.
In \refeq{first}, $\lambda_{mat}$, $\rho_{mat}$, and $c_{mat}$ denote the heat conductivity, mass density, and heat capacity of concrete, respectively, and $\frac{\partial Q_h}{\partial t}$ the rate of hydration heat released, see~\refsec{Heat_development_during_cement_hydration}.
In \refeq{second}, $D_H$ represents the diffusivity coefficient and $\kappa$ the hygrothermic coefficient ensuring a weak coupling between temperature and humidity, see Appendix~\ref{sec:derivHumidity}.
The sink term $\frac{\partial H_s}{\partial t}$ accounts for the self-desiccation of the material due to hydration.

While~\refeq{governing_eq_strong-form} describe a general transient heat and humidity conduction over a domain and thus form a backbone of the computational model, we will now introduce three aspects to focus its broad applicability to early-age hydrating concrete.
In particular, we focus on i) the rate of hydration heat, which acts as an internal heat source in the concrete structure, followed by ii) the description of an analytical model to characterize microstructural development, and finally arriving at iii) the description of thermophysical properties of concrete and their link to microstructural evolution.

\subsection{Internal heat source from cement hydration}
\label{sec:Heat_development_during_cement_hydration}
Upon contact with water, cement particles undergo hydration reactions resulting in a measurable heat release.
Isothermal calorimetry permits continuous measurements of the cumulative heat of hydration $Q_h(t,T_{\infty},H_{\infty})$ in [J/kg] released for a given time $t$~[s]. Heat production depends on ambient temperature $T_{\infty}$~[$^{\circ}$C] and relative humidity $H_{\infty}$~[-] history, and can be expressed as
\begin{linenomath}
	\begin{eqnarray}
		Q_h(t,T_{\infty},H_{\infty}) = \alpha \, Q_{\text{tot}}(t\to \infty), \quad \alpha \in \langle0,1\rangle,
		\label{eq:Qh}
	\end{eqnarray}
\end{linenomath}
where $\alpha$~[-] denotes the achieved state of microstructural development, described here as the degree of cement hydration, and $Q_{\text{tot}}(t\to \infty)$ in [J/kg] denotes the latent hydration heat of cement.
The latter is unique for each cement as it depends on its mineralogical composition. 
It can, however, be estimated as a weighted average of enthalpies of the main clinker constituents at complete hydration~\cite[p.\ 232]{Taylor:1997}.

Focusing on the rate of hydration heat originating from cement dissolution $\frac{\partial Q_h}{\partial t}$ in [J/(s~m$^3$)], we find 
\begin{linenomath}
	\begin{eqnarray}
		\frac{\partial Q_h}{\partial t} = \frac{\partial \alpha}{\partial t} \, Q_{\text{tot}}(t\to \infty),
		\label{eq:dQh}
	\end{eqnarray}
\end{linenomath}
where the rate of microstructural development 
\begin{linenomath}
	\begin{eqnarray}
		\frac{\partial \alpha}{\partial t} = \tilde{A}(t, T_{\infty}, H_{\infty}) \, ,
		\label{eq:dalpha}
	\end{eqnarray}
\end{linenomath}
is described by $\tilde{A}(t, T_{\infty}, H_{\infty})$ in [s$^{-1}$], the chemical affinity, and is defined as the change of microstructural development in time.
This term was introduced by Ulm and Coussy~\cite{Ulm:1995} and conjugates the gradient of free enthalpy between the reactants and products of the chemical reactions occurring during hydration~\cite{Cervera:1999, Gawin:2006}.
Consequently, $\tilde{A}(t, T_{\infty}, H_{\infty})$ is a time-, temperature-, and humidity-dependent quantity.
The time specification will be replaced by $\alpha$ in $\tilde{A}(\alpha, T_{\infty}, H_{\infty})$, recalling that $\alpha$ defines the \textit{resultant} state of the microstructural evolution, regardless of when and under which conditions it was achieved.

The rate of hydration heat defined by~\refeq{dQh} physically represents an internal heat source within the concrete structure.
This term, therefore, appears in the governing equation for transient heat conduction, see~\refeq{first}.
\refeq{dalpha} describing microstructural development serves as the final governing equation, together with~\refeq{governing_eq_strong-form}.

\subsection{Affinity-based hydration model}\label{sec:Affinity-based_hydration_model}
Experimental determination of the hydration heat rate for a given cement type, mixture composition, and variable temperature requires repeated measurements, which are time-intensive and, hence, impractical for modeling purposes.
In addition, fluctuations in curing temperature and/or ambient humidity affect the reaction kinetics of cement hydration.
To account for both phenomena separately, we use dimensionless correction factors $\beta_\text{T}$ and $\beta_\text{H}$~\cite{Gawin:2006}. 
These will be used to scale the cement hydration kinetics from a reference state defined by $\tilde{A}_\text{ref}$ as follows
\begin{linenomath}
	\begin{eqnarray}
		\tilde{A}(\alpha, T_{\infty},H_{\infty}) = \tilde{A}_{\text{ref}}(\alpha) \, \beta_\text{T}(T_{\infty}) \, \beta_\text{H}(H_{\infty}) \, .
		\label{eq:Atld}
	\end{eqnarray}
\end{linenomath}
Both scaling factors are discussed separately below.

For a computational implementation, we consider a phenomenological affinity-based hydration model that provides an analytical relationship between microstructural development in a hydrating paste and concurrent hydration heat.
This model was initially formulated by Cervera et al.~\cite{Cervera:1999} and later modified by Jendele et al.~\cite{Jendele:2014}.
We adopt the latter formulation, which defines $\tilde{A}_{\text{ref}}$ as
\begin{linenomath}
	\begin{eqnarray}
		\tilde{A}_{\text{ref}} = B_1 \left( \frac{B_2}{\alpha_{\infty}} + \alpha  \right) (\alpha_{\infty} - \alpha) \exp \left( -\eta \, \frac{\alpha}{\alpha_{\infty}}  \right),
		\label{eq:affinity}
	\end{eqnarray}
\end{linenomath}
where $\tilde{A}_{\text{ref}}$ needs to be determined at a reference temperature, $B_1$~in [s$^{-1}$], $B_2$~[-] and $\eta$~[-] are best-fit parameters characterizing the kinetics of cement hydration and $\alpha_{\infty}$ represents the maximum-reachable degree of hydration of the cement paste.
The initial three parameters need to be first calibrated for the specific cement.
In this paper, we use isothermal calorimetry measurements for calibration, but if such results are not available, suitable hydration models can be used instead (\textit{e.g.},\ CEMHYD3D~\cite{Bentz:1997, Hlobil:2022cementPSD}).
The term $\alpha_{\infty}$ then depends on the fresh mix design and curing conditions~\cite{Hansen:1986}, see Appendix~\ref{appendix:Calibration_affinity-model} for details.

As cement hydration is a thermally activated process, an Arrhenius-type equation can be used to correct $\tilde{A}_{\text{ref}}$ for temperature fluctuation in~\refeq{Atld}. 
Consequently, we consider the coefficient $\beta_\text{T}$~[-] as
\begin{linenomath}
	\begin{eqnarray}
		\beta_\text{T}(T) = \exp \left[ \frac{E_a}{R} \left( \frac{1}{T_{\infty}} - \frac{1}{T_{mat}}\right)  \right],
		\label{eq:beta_T}
	\end{eqnarray}
\end{linenomath}
where $T_{\infty}=293.15$~K represents the reference ambient temperature for which $\tilde{A}_{\text{ref}}$ is determined, $R=8.31441$~J/(mol~K) is the universal gas constant, and $E_a=38.3$~kJ/mol is the apparent activation energy based on measurements from Kada-Benameur et al.~\cite{Kada:2000}.
The latter quantity is a mixture- and temperature-dependent parameter that characterizes the binder's integrated and simultaneous chemical reactions.

Cement hydration encompasses a series of chemical processes, all requiring a constant supply of liquid water from the capillary pore network.
This facilitates the transport of ionic species and the subsequent formation of reaction products. 
However, the rate of hydration progressively decreases with the dropping of pore relative humidity and eventually stops upon reaching~$\text{H} \approx 0.80$~\cite{Jensen:1998}.
As a remedy, we recall the coefficient $\beta_\text{H}$~[-] coined in~\cite{Bazant:1972MaS} into~\refeq{Atld}, which accounts for the reduction of hydration rate for decreasing the capillary humidity as
\begin{linenomath}
	\begin{eqnarray}
		\beta_\text{H}(H) = \frac{1}{1+ {\left( a-a \, H_{mat} \right)}^4 } ,
		\label{eq:beta_H}
	\end{eqnarray}
\end{linenomath}
where $a=7.5$~[-] is a material parameter introduced in~\cite{Bazant:1972MaS}.

\subsection{Thermophysical properties of concrete constituents}
The parameters $\rho_{mat}$, $c_{mat}(\alpha)$, and $\lambda_{mat}(\alpha)$ postulated in~\refeq{governing_eq_strong-form} characterize the physical and thermal properties of concrete.
For the development of the model, we consider $\rho_{mat}$ as constant, unaffected by ongoing hydration or water evaporation from the exposed structural surface.
However, $c_{mat}(\alpha)$ and $\lambda_{mat}(\alpha)$ evolve in time, reflecting the temperature- and humidity-dependent microstructural development.

The specific heat capacity of a material, $c_{mat}$, is defined as the amount of heat necessary to change the temperature of the material by one degree, normalized per mass of the material.
An expression for the heat capacity of hydrating cement paste, $c_{paste}(\alpha)$, should account for the initial paste composition given by the water-to-cement mass ratio $w/c$, ongoing microstructural development, and curing conditions~\cite{Bentz:2007}. 
We here consider 
\begin{linenomath}
	\begin{eqnarray}
		c_{paste}(\alpha) &=& \frac{c_{water} \, w/c + c_{cement}}{1+w/c} \, (1-0.26\, (1-\exp(-2.9\,\alpha)) ),
		\label{eq:heat_capacity}
	\end{eqnarray}
\end{linenomath}
where $c_{water}=4.18$~J/(g K) and $c_{cement}=0.75$~J/(g K) stand for the heat capacities of water~\cite{Holman:1981} and anhydrous cement~\cite{Todd:1951} measured at room temperature, respectively.

The heat capacity of concrete, $c_{concrete}$, analogously depends on the properties of both cement paste and aggregates and can be estimated~\cite{Bentz:2007} using a simple rule of mixtures as 
\begin{linenomath}
	\begin{eqnarray}
	c_{concrete} = m_{f,aggr} \, c_{aggr} + m_{f,paste} \, c_{paste}(\alpha), 
	\label{eq:heat_capacity_concrete}
	\end{eqnarray}
\end{linenomath}
where $m_{f,phase}$ stands for a mass fraction of the given phase, and $c_{aggr}= 0.8$~J/(g K)~\cite{Glinicki:2015}.

The heat conductivity $\lambda_{mat}$ is defined as the ratio of heat flux to the temperature gradient~\cite{Neville:1997}. 
For cement paste, $\lambda_{paste} \approx 0.9-1.1$~W/(m K) remains virtually unchanged throughout hydration regardless of the mix composition and curing conditions.
This has been experimentally verified in~\cite{Bentz:2007, Mounanga:2004} and computationally corroborated in~\cite{Qomi:2015, Honorio:2018}.
A virtually constant $\lambda_{paste} = 1.0$~W/(m K) can be anticipated based on a comparison of heat conductivities of constituting solid phases in the paste and their envisioned relative proportions throughout hydration ($\lambda_{clinker}=1.55$~\cite{Bentz:2007}, $\lambda_{C-S-H}=0.98$, $\lambda_{CH}=1.32$ (both from~\cite{Qomi:2015}), $\lambda_{water}=0.604$~W/(m~K)~\cite{Bentz:2007}). 
While a minor decrease of $\lambda_{paste}$ with $w/c$ ratio was observed in~\cite{Mounanga:2003phd}, the differences may be deemed practically negligible for the considered range of $w/c \in 0.25-0.4$~[-].

The initial concrete conductivity $\lambda_{concrete}$ is largely governed by the relative proportions of aggregates and paste in the mixture.
Bentz~\cite{Bentz:2007} proposed an estimate to evaluate the initial $\lambda_{concrete}^{init}$ of fresh concrete using a Hashin-Shtrikman homogenization scheme, with $\lambda_{concrete}$ as an arithmetic mean from the upper and lower bounds given as 
\begin{linenomath}
	\begin{eqnarray}
		\lambda_{concrete}^{upper} &=& \lambda_{aggr} + \frac{\phi_{paste}}{\frac{1}{\lambda_{paste}-\lambda_{aggr}} + \frac{\phi_{aggr}}{3\, \lambda_{aggr}}},\\
		\lambda_{concrete}^{lower} &=& \lambda_{paste} + \frac{\phi_{aggr}}{\frac{1}{\lambda_{aggr}-\lambda_{paste}} + \frac{\phi_{paste}}{3\, \lambda_{paste}}},\\
		\lambda_{concrete}^{init} &=& \frac{\lambda_{concrete}^{upper}+\lambda_{concrete}^{lower}}{2},
		\label{eq:HS_bounds_concrete}
	\end{eqnarray}
\end{linenomath}
where $\lambda$ and $\phi$ denote the heat conductivity and volume fraction of a given component (paste or aggregates); here, we consider $\lambda_{aggr}= 3.0$~W/(m K)~\cite{Fei:2022}.
Note that~\refeq{HS_bounds_concrete} assumes that $\lambda_{aggr} > \lambda_{paste}$, which is valid for most siliceous and calcitic aggregates but excludes lightweight aggregates.

Experimental measurements by Chen et al.~\cite{Chen:2022} indicate that concrete heat conductivity further evolves with ongoing cement hydration and is strongly affected by curing conditions.
Based on their experimental results, we consider the evolution of $\lambda_{concrete}$ under sealed conditions as
\begin{linenomath}
	\begin{eqnarray}
		\lambda_{concrete}^{sealed}(\alpha) &=& \lambda_{concrete}^{init} \, (1+0.22 \, \alpha) \, .
		\label{eq:lambda_evo}
	\end{eqnarray}
\end{linenomath}
Note that while the experimentally measured concrete conductivity increases under sealed conditions in time, the opposite trend was observed for saturated conditions.
Chen et al.~\cite{Chen:2022} attribute this observation to the differences in the microstructural development during hydration, particularly in the presence of water within the overall pore network, from capillary pores to the interfacial transition zone around aggregates.
During sealed curing, the material experiences self-desiccation as capillary pores are progressively emptied.
As the solid phase fraction increases, the moisture content decreases, so the overall conductivity of the material becomes proportional to the conductivity of the solid skeleton matrix.
In comparison, water remains omnipresent within the pore network during the saturated curing, thus effectively lowering the overall conductivity.
However, the only former curing regime applies in realistic conditions for extruded concrete structures.

\section{Numerical implementation}
Surface drying of hydrating concrete can be described by a set of governing Equations given by~\refeq{governing_eq_strong-form} and (\ref{eq:dalpha}).
To find an approximate solution, we employ the Finite Element Method implemented in the open-source FEniCS package~\cite{Logg:2012}.
First, we spatially discretize a rectangular domain representing a 2D slice of the extruded concrete structure.
To this end, we consider a full structural cross-section, grouping individual filaments into a materially homogeneous domain as shown in~\reffig{mmodel}(b). 
Next, we resolve the coupled transient conduction phenomena over the entire domain by discretizing the time derivatives using a finite difference approximation with a backward difference quotient, yielding an implicit Euler method.
We solve for the sought $\alpha$, $T$, and $H$ fields, which act as field variables.

Solving the set of governing equations given in~\refeq{governing_eq_strong-form} and (\ref{eq:dalpha}) requires suitable initial and boundary conditions, reflecting the interaction of the extruded structure with the surrounding environment.
The initial condition (IC) denotes the starting value of the degree of hydration, temperature, and humidity fields, assumed herein as $\alpha(x,0)=0$~[-], $T_{mat}(x,0)=20$\degC, and $H_{mat}(x,0)=1.0$~[-] over the entire domain.
We model the heat exchange between the structural surface and its environment with Robin's boundary condition (BC) as ${\overline{q}_T}(t) = h_T(T_{mat}(x,t) - T_{\infty})$, where $h_T$ the heat transfer coefficient in [W/(m$^2$ K)], $T_{mat}(x,t)$ the surface temperature of the material, and $T_{\infty}=20$\degC is the remote (ambient) temperature.
An analogous form of Robin BC is also used for humidity transport with ${\overline{q}_H}(t) = h_H(H_{mat}(x,t) - H_{\infty})$ with $h_H$ as the hygric exchange coefficient in [m/s], $H_{mat}(x,t)$ the surface humidity of the material, and $H_\infty=0.60$~[--] as the remote (ambient) humidity.
The listed values represent the default settings used for model evaluation and are applied unless noted otherwise.

\section{Model verification}\label{sec:model_verif}
\subsection{Calibration of the affinity-based hydration model}\label{sec:verif_hydration}
Here, we demonstrate the affinity-based model's descriptive capability for assessing the hydration kinetics of various cement pastes.
To this end, we consider experimental results~\cite{Hlobil:2022cementPSD} from isothermal calorimetry tests of pastes, which differ in cement type and $w/c$ ratio used.
Notably, all considered pastes were free from chemical admixtures that could alter the hydration kinetics.
We plot the experimental results, interpreted by $\tilde{A}_\text{ref}$ as a function of the achieved microstructural development $\alpha$, and fit the analytical form from~\refeq{affinity} using a damped least-squares algorithm.
This method provides the sought set of parameters $B_{\text{1}}$, $B_{\text{2}}$, and $\eta$, summarized in Appendix~\ref{appendix:Calibration_affinity-model}, which are independent for each paste tested and characterize their hydration kinetics.

\begin{figure}[htb!]
		\centering
		\includegraphics[scale=1]{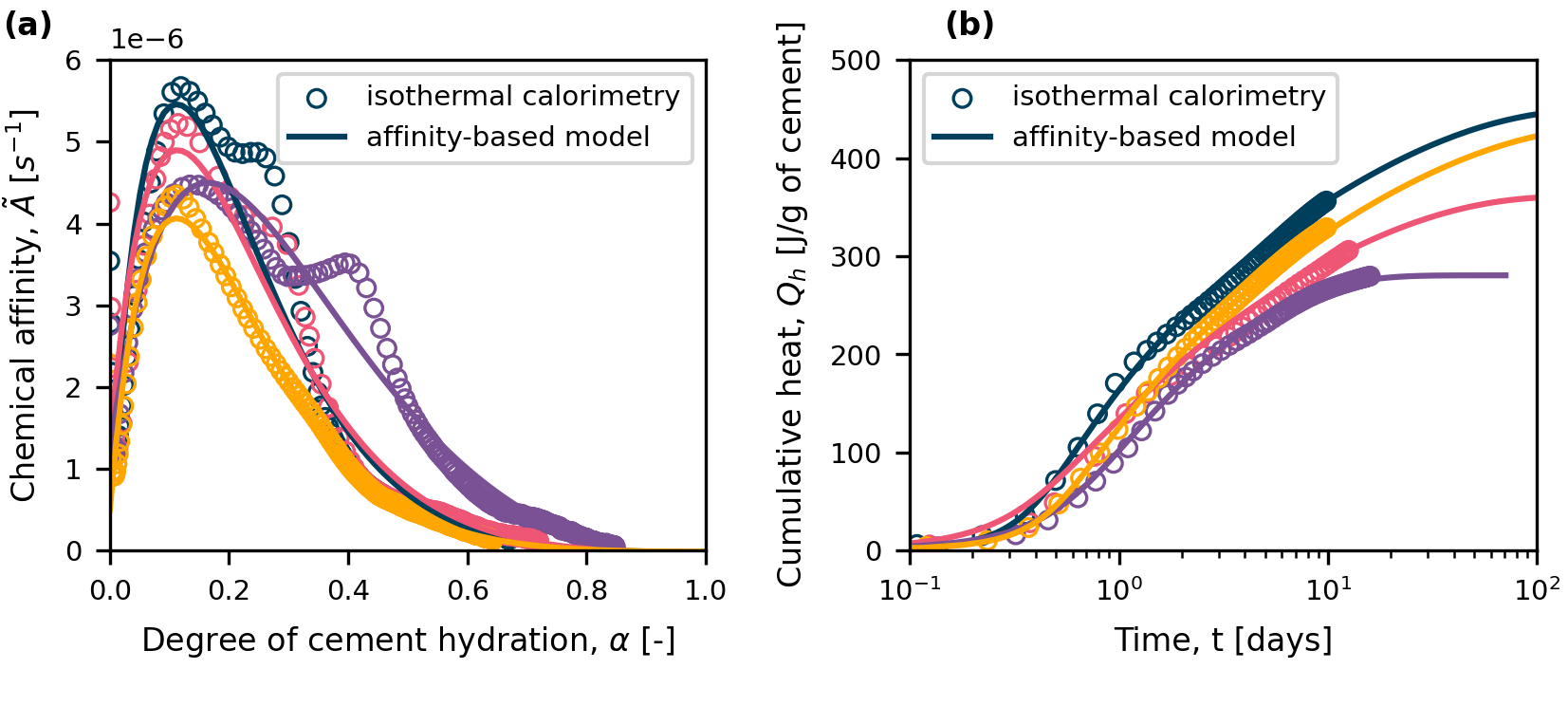}
		\caption{Application of the analytical affinity-based hydration model to quantify cement hydration. (a) Calibration of the model to experimental data from isothermal calorimetry. (b) Quantitative evaluation of heat release after model calibration.}
	\label{fig:affinity-model-predictions}
\end{figure}

The affinity-based hydration model captures the main hydration peak and subsequent decrease of reaction kinetics with acceptable accuracy, as shown in~\reffig{affinity-model-predictions}(a). 
We then use~\refeq{Qh} to calculate the time-evolution of cumulative heat for each paste and compare the calculated results to experiments in~\reffig{affinity-model-predictions}(b).
The analytical model shows a satisfactory agreement with the measured data regardless of the cement paste composition.
Thus, it provides an analytical link between microstructural formation and corresponding heat release.
This achievement enables us to evaluate the temperature increase in concrete structures caused by cement hydration, as discussed next.
\FloatBarrier

\subsection{Temperature increase in hardening concrete}\label{sec:verif_temperature}
\begin{figure}[htb!]
		\centering
		\includegraphics[scale=1]{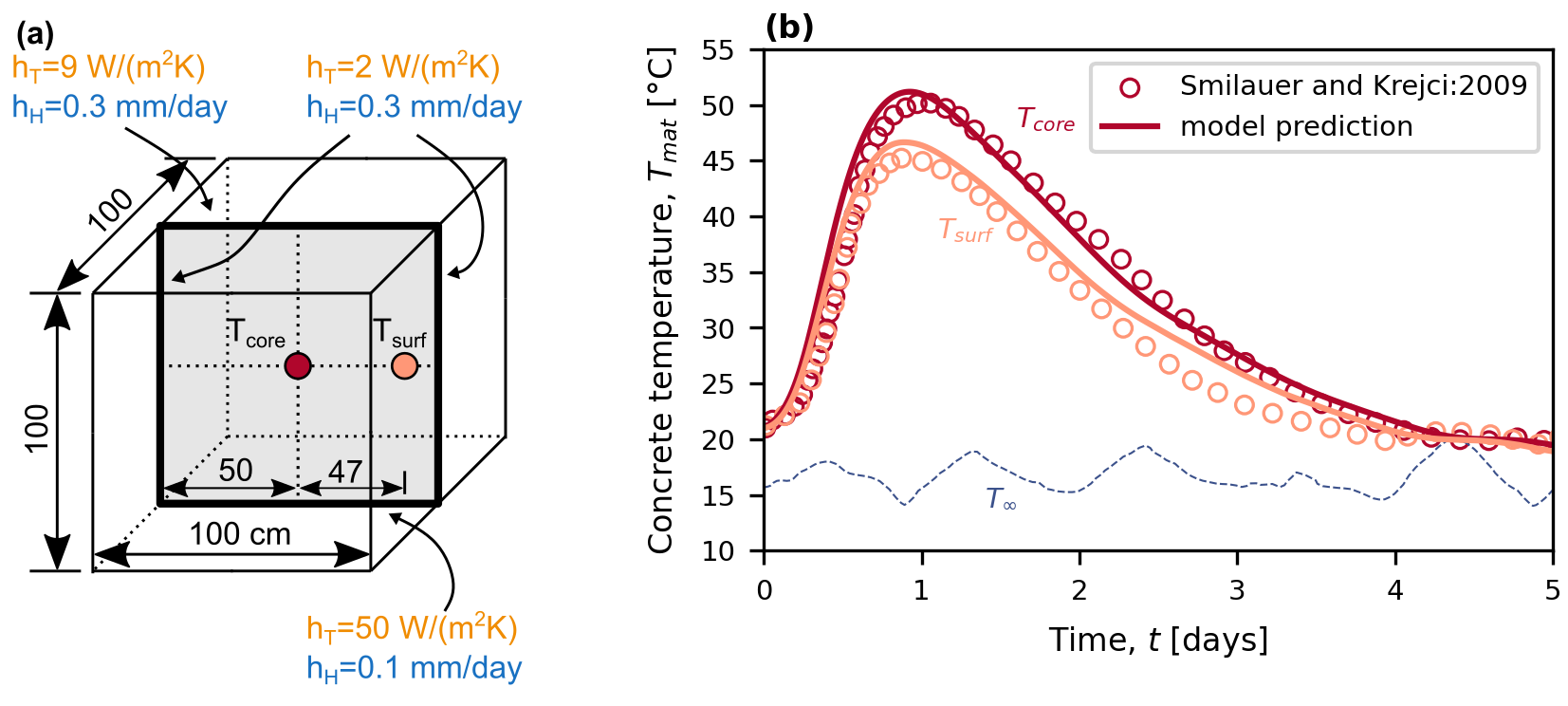}
		\caption{Simulation of temperature increase in a hydrating concrete block. (a) Modeled 2D cross-section and considered heat and humidity transfer coefficients. (b) Computed results shown as solid lines and circles indicate measurements from~\cite{Smilauer:09}. The dashed line shows recorded ambient temperature fluctuation, which was imposed as a boundary condition.}
		\label{fig:verif_Smilauer-cube}
\end{figure}

We analyze the effect of hydration heat on the temperature increase within a massive concrete block, as recorded by \v{S}milauer and Krej\v{c}\'{i}~\cite{Smilauer:09}.
In this experiment, a cubic block with a side length of 1~m was cast from self-compacting concrete and kept in a protective shelter, shielding it from direct sun radiation and rain.
Monitoring gauges were embedded into the block, recording the temperature evolution at the core and 3~cm from its outer surface, as well as ambient temperature, see~\reffig{verif_Smilauer-cube}(a).
The block was separated from a solid concrete foundation by a 2 mm thick impregnated paper.
A 15-mm-thick plywood formwork protected its sides, whereas the top surface was exposed to the ambient environment.

We simulate the experiment as a coupled temperature and humidity problem within a 2D cross-section, representing a central slice of the specimen as shown in~\reffig{verif_Smilauer-cube}(a).
We considered the kinetics of a generic cement with a common mineralogical composition, see Appendix~\ref{appendix:Params4modelEvaluation}, since the reported data in~\cite{Smilauer:09} lacked the necessary inputs for precise calibration of the hydration model.
We prescribe a variable heat and humidity flow to each block boundary, as shown in~\reffig{verif_Smilauer-cube}(a), to account for the fluctuating $T_{\infty}$ but suppose $H_{\infty}=0.6$~[-] since it was not reported.

Computed results show a satisfying prediction of temperature evolution, correctly capturing the main peak, as well as an acceptable cooling regime, see~\reffig{verif_Smilauer-cube}(b).
The discrepancy in surface temperature observed after 2 days from casting can probably be attributed to the removal of formwork, which would affect the chosen value of the prescribed transfer coefficients.
However, such an event was not explicitly documented in~\cite{Smilauer:09}, and therefore, we kept $h_T$ and $h_H$ constant throughout the simulation.
Nonetheless, we consider the simulation results satisfactory and emphasize the correctly calculated peak temperature and initial cooling stage.
\FloatBarrier

We note that the computed results were based solely on empirical knowledge of the specimen (geometry, material composition) and its interaction with the environment (surface treatment and ambient temperature).
We chose the values for the transfer coefficients ($h_T$ and $h_H$) and ambient humidity $H_{\infty}$ in line with common considerations for modeling transport phenomena in concrete in a protected environment~\cite{Azenha:2021, Smilauer:2019, Smilauer:09, Havlasek:2021}.
While a mathematical optimization of the used values could provide an even closer agreement with the measured data, such analysis is beyond the scope of this paper.

\subsection{Drying of concrete elements}\label{sec:verif_drying}
We verify the suitability of the presented model for predicting concrete drying by analyzing experimental measurements carried out by Vinker and V\'{i}tek~\cite{Vinkler:2016a, Vinkler:2019}.
Their data contains measured short-term self-desiccation of a standardized concrete cube over 2 months and long-term relative humidity measurements in a massive concrete element over 3 years.
We will use our model to simulate the humidity evolution in each specimen separately to account for the differences in specimen size, curing regime, and at different depths from the specimen surface and compare with the experimental data.

\begin{figure}[htb!]
    \centering
    \includegraphics[scale=1]{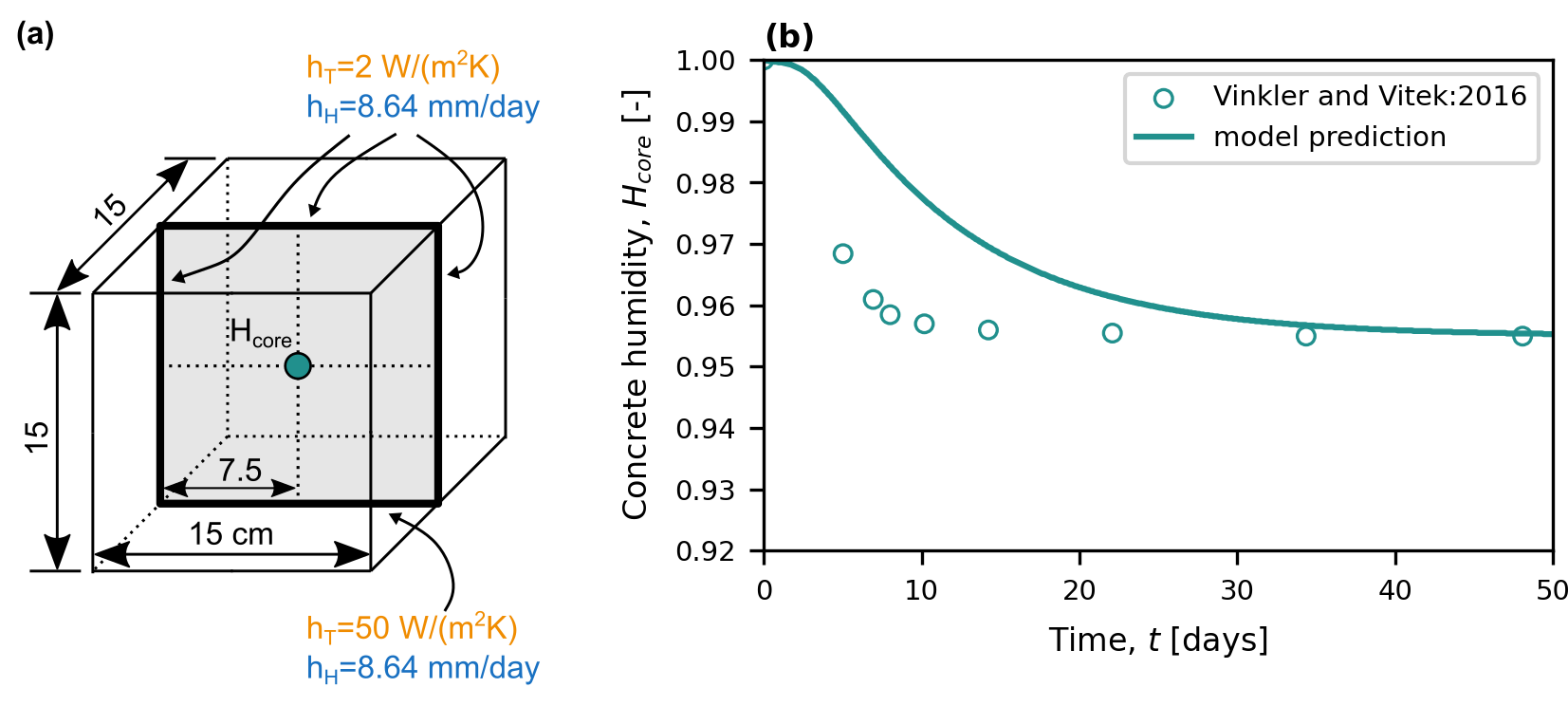}
    \caption{Simulation of self-desiccation within a standardized concrete cube. (a) Modeled 2D cross-section and considered heat and humidity transfer coefficients. (b) Computed results shown as a solid line and circles indicate measurements from~\cite{Vinkler:2016a, Vinkler:2019}.}
	\label{fig:concrete_drying_cube}
\end{figure}

The first specimen assessed was a standardized concrete cube with a side of 150 mm that was wrapped in foil and sealed in a protected environment upon demolding.
The internal humidity decrease from self-desiccation was recorded at the cube's core at irregular intervals over two months; see points in~\reffig{concrete_drying_cube}(a).
To model this experiment, we considered a coupled temperature and humidity simulation over a 2D domain representing the central slice of the cube.
We prescribed a constant ambient humidity as $H_{\infty}=0.955~[-]$ to account for the sealed conditions and transport coefficients as shown in~\reffig{concrete_drying_cube}(a), as well as the cement kinetics parameters summarized in~Appendix~\ref{appendix:Params4modelEvaluation}.
\FloatBarrier

\begin{figure}[htb!]
    \centering
    \includegraphics[scale=1]{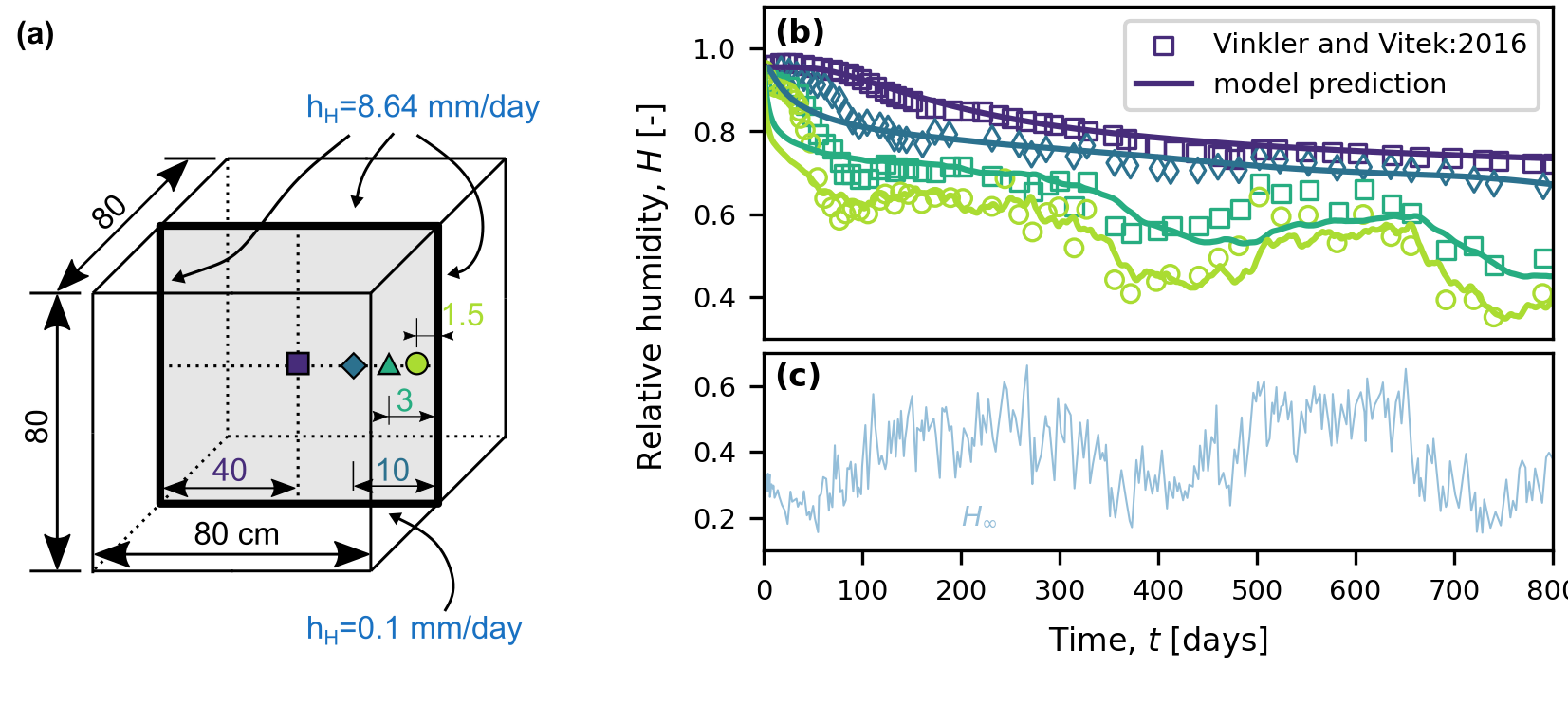}
    \caption{Drying simulation of a massive concrete element over 800 days. 
    (a) Modeled 2D cross-section and considered humidity transfer coefficients. (b) Computed results are shown as solid lines, and points indicate measurements from~\cite{Vinkler:2016a, Vinkler:2019}. (c) Recorded ambient humidity fluctuation, which was imposed as a boundary condition.}
	\label{fig:concrete_drying_element}
\end{figure}

Simulation results capture the drop in internal relative humidity at the cube core with acceptable accuracy for engineering purposes, differing only by single units of \% from experiments.
We attribute the measured rapid drop in humidity down to $H_{mat} \approx 0.96$~[-] during the first week to specimen demolding and subsequent short-term exposure to low ambient humidity before wrapping; refer to early measured values of $H_{\infty}$ in~\reffig{concrete_drying_element}(b).
We will show in \refsec{ambient_conditions} that even a short exposure to dry ambient conditions leads to a rapid acceleration of material drying.
Deplorably, the demolding event is not explicitly documented in~\cite{Vinkler:2016a, Vinkler:2019}, yet it plays a crucial role in $H_{mat}$ development.
As a conclusion, exposure to a sub-saturated ambient environment accelerates the humidity exchange with the environment, leading to material drying.

The second experiment involved continuous humidity measurements within a massive concrete cube with a side length of 800 mm~\cite{Vinkler:2016a, Vinkler:2019}.
The material composition was identical to the standardized cube described above, but the element's surface was kept uncovered and thus exposed to a variable $H_{\infty}$; see~\reffig{concrete_drying_element}(C).
Internal humidity measurements were continuously recorded over 800 days at different depths within the specimen.
We simulated the experimental setup by considering a 2D planar cross-section representing the central slice of the element.
Given the measurement duration of over 800 days, any temperature-induced hydration effects and microstructural development in such material become negligible after the first few weeks.
Consequently, we carried out only a transient humidity simulation.
To simulate the drying of a mature concrete element, we set the initial material humidity as $H_{mat}=0.955$~[-] in line with the self-desiccation measurements of concrete reported above.
We then applied a fluctuating $H_{\infty}$ as the BC on the top and sides of the model; the rest of the numerical inputs are summarized in~\reffig{concrete_drying_element}(a).
\FloatBarrier

Simulation results shown in~\reffig{concrete_drying_element}(b) indicate that the fluctuation of ambient humidity $H_{\infty}$ affects the internal humidity $H_{mat}$ within the concrete element.
At shallow depths from the surface (here depths of 15 and 30 mm), $H_{mat}$ evolution mirrors the fluctuation of $H_{\infty}$.
For depths below 100~mm, the influence of $H_{\infty}$ on $H_{mat}$ vanishes, as experimentally measured in~\cite{Vinkler:2016a, Vinkler:2019} and obtained computationally herein.
As a general observation, the continuous drying of the entire specimen follows from the long-term exposure to a sub-saturated ambient environment, as accurately simulated by the computational model.
This successful model verification serves as a stepping stone to quantify a surface humidity threshold potentially leading to cold joint quantification, as described below.

\section{Surface humidity threshold for cold joint formation}\label{sec:cj_prediction}
While concrete surfaces naturally dry over time, we focus on identifying a humidity threshold that would indicate a potential threat of cold joint formation.
To this end, we consider the experimental data measured by Keita et al.~\cite{Keita:2019}, who quantified the drying rate of fresh concrete under controlled conditions.
In their work, they exposed freshly cast specimens to a forced airflow in a climatic chamber and recorded the exposure time.
Measured results indicate that surface drying after 10 minutes of exposure had detrimental consequences on the material's mechanical performance.

\begin{figure}[htb!]
		\centering
		\includegraphics[scale=1]{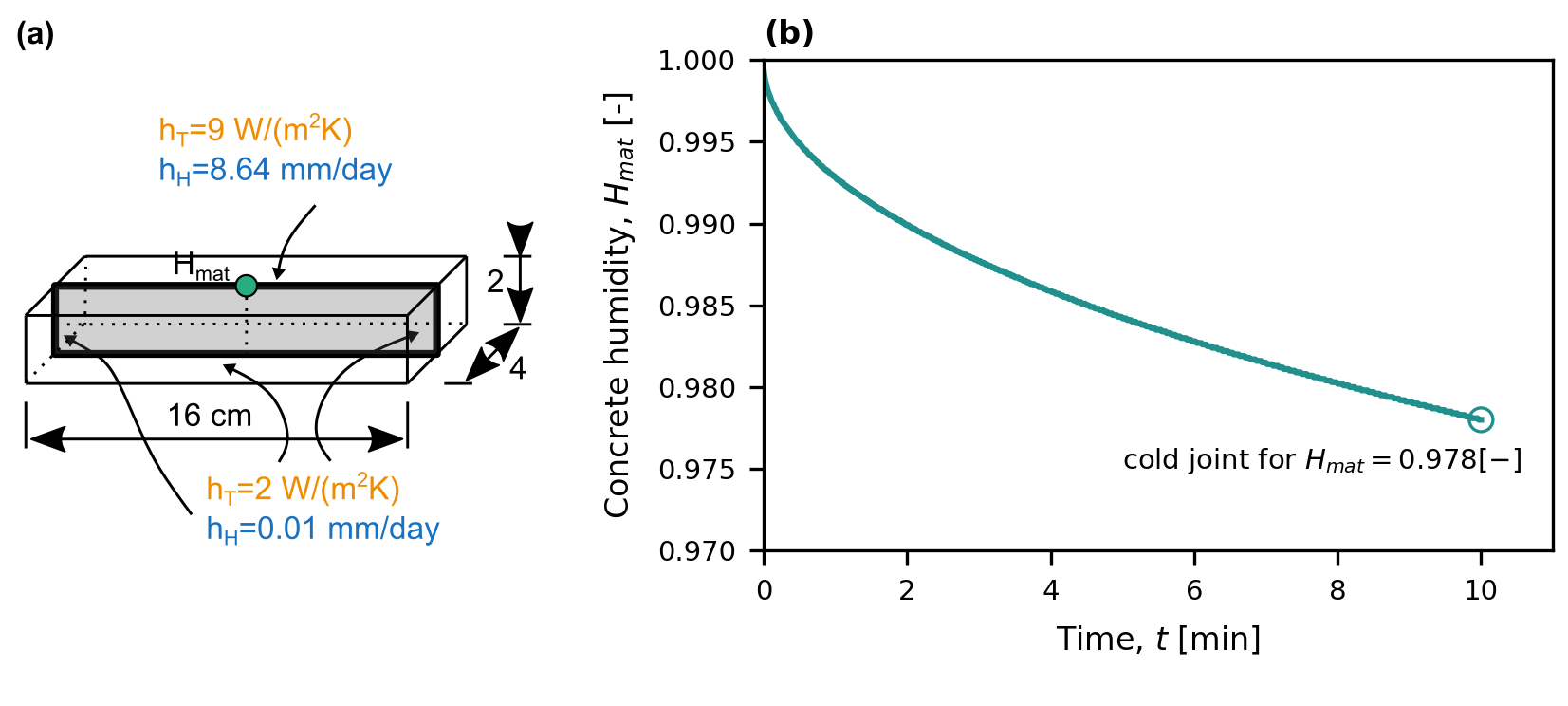}
		\caption{Determination of surface humidity threshold leading to a cold joint formation. (a) Modeled 2D cross-section and considered heat and humidity transfer coefficients. (b) The computed surface humidity as it evolves over time.}
		\label{fig:cold-joint-sim}
\end{figure}

We simulate the aforementioned drying experiment from Keita et al.~\cite{Keita:2019} to determine a surface humidity value leading to a potential cold joint formation.
For model evaluation, we consider a central 2D slice of the tested specimen, see~\reffig{cold-joint-sim}(a), and perform a coupled temperature and humidity simulation with cement kinetics parameters summarized in~Appendix~\ref{appendix:Params4modelEvaluation}.
The specimen geometry and considered BC are shown in~\reffig{cold-joint-sim}(a).
The computed relative humidity at the top specimen surface after 10 minutes of forced drying amounted to $H_{mat} = 0.978$~[-], see result in~\reffig{cold-joint-sim}(b).
From now on, we will consider this value as a humidity threshold in~\refsec{cj_formation} to consistently compare the factors affecting surface drying and assess their severity under given exposure conditions.
\FloatBarrier

Given the lack of experimental data on surface humidity measurements leading to cold joint formation, we simulate the well-defined drying experiment from~\cite{Keita:2019} and obtain a calculated humidity threshold value.
We acknowledge that our assessment will rely on a single data point and thus cannot be proclaimed a universally applicable criterion for cold joint formation.
Nonetheless, it allows for a qualitative assessment of the various input factors.
More experimental data on surface humidity evolution should be collected before any concluding recommendation can be made.

\section{Factors affecting cold joint formation}\label{sec:cj_formation}
\begin{figure}[htb!]
    \centering
    \includegraphics[scale=1]{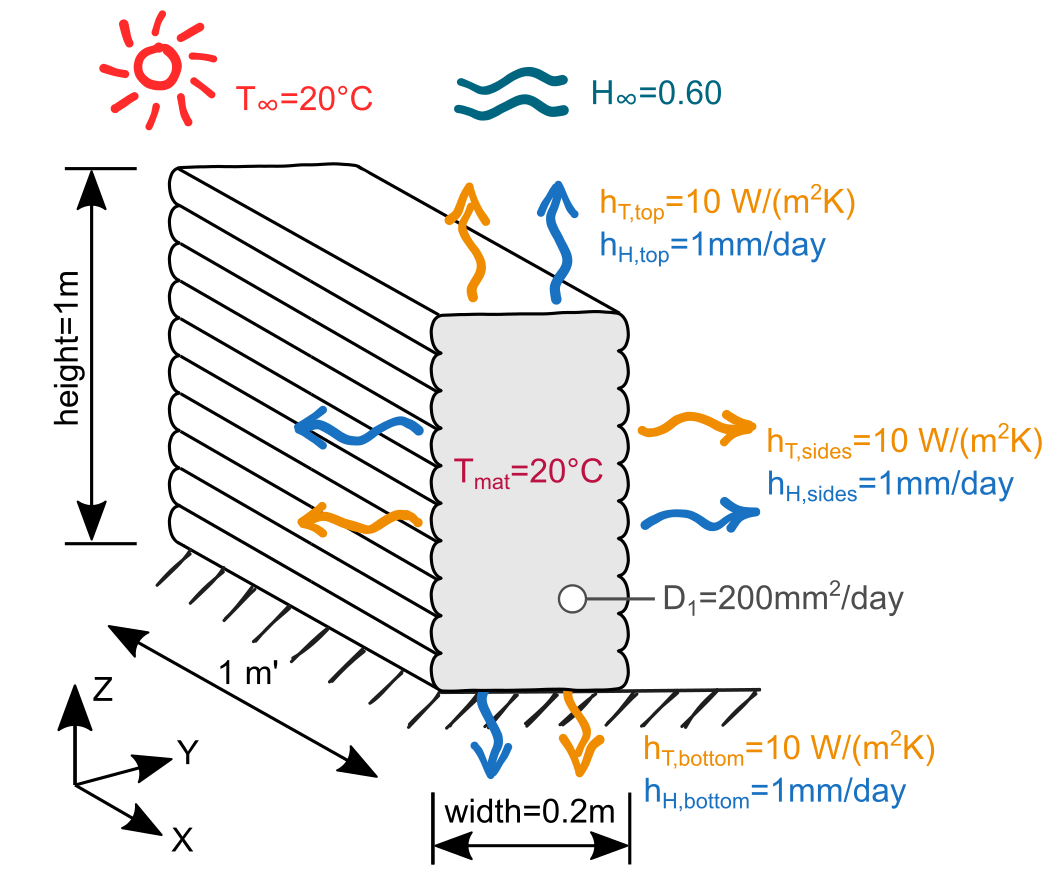}
	\caption{A fictitious concrete structure used to assess the drying severity for a given structure’s geometry, its interaction with the environment, and ambient conditions.}
	\label{fig:structure_modelInputs}
\end{figure}

Various factors potentially influence the surface drying of extruded concrete.
To evaluate the risk of cold joint formation, we first envision a fictitious concrete structure with a realistic geometry, extruded within a common ambient environment, as schematically shown in~\reffig{structure_modelInputs}.
Next, we perform a coupled heat and humidity transport simulation and evaluate the time $t_{cj}$ necessary for the humidity on the top surface of the structure to reach the threshold, potentially leading to a cold joint formation.
For the listed combination of inputs, the computed time amounts to $t_{cj} = 501$~min.
While such duration may seem excessive and far from threatening in a typical use case, a combination of the factors listed below can drastically reduce the time to units of minutes, raising concerns in all but the fastest extrusions of structures.
\FloatBarrier

To assess the severity of drying for each factor separately, from the state of material development to structural geometry over to the influence of the ambient environment, we sequentially vary the value for each factor within experimentally achievable bounds and compare the resulting times.
To this end, we introduce
\begin{linenomath}
	\begin{eqnarray}
		\chi= \frac{t_i}{t_{cj}},
		\label{eq:timeratio}
	\end{eqnarray}
\end{linenomath}
with $t_i$ denoting the simulated time to reach the threshold humidity for a given variation in the input factor. 
\refeq{timeratio} provides a relative scaling factor that indicates either accelerated ($\chi<1$) or hindered ($\chi>1$) time for cold joint formation and allows the assessment of the severity of surface drying.
In the following subsections, we discuss each factor's influence and provide recommendations for possible cold joint mitigation.

\subsection{Ambient conditions}\label{sec:ambient_conditions}
\begin{figure}[htb!]
    \centering
    \includegraphics[scale=1]{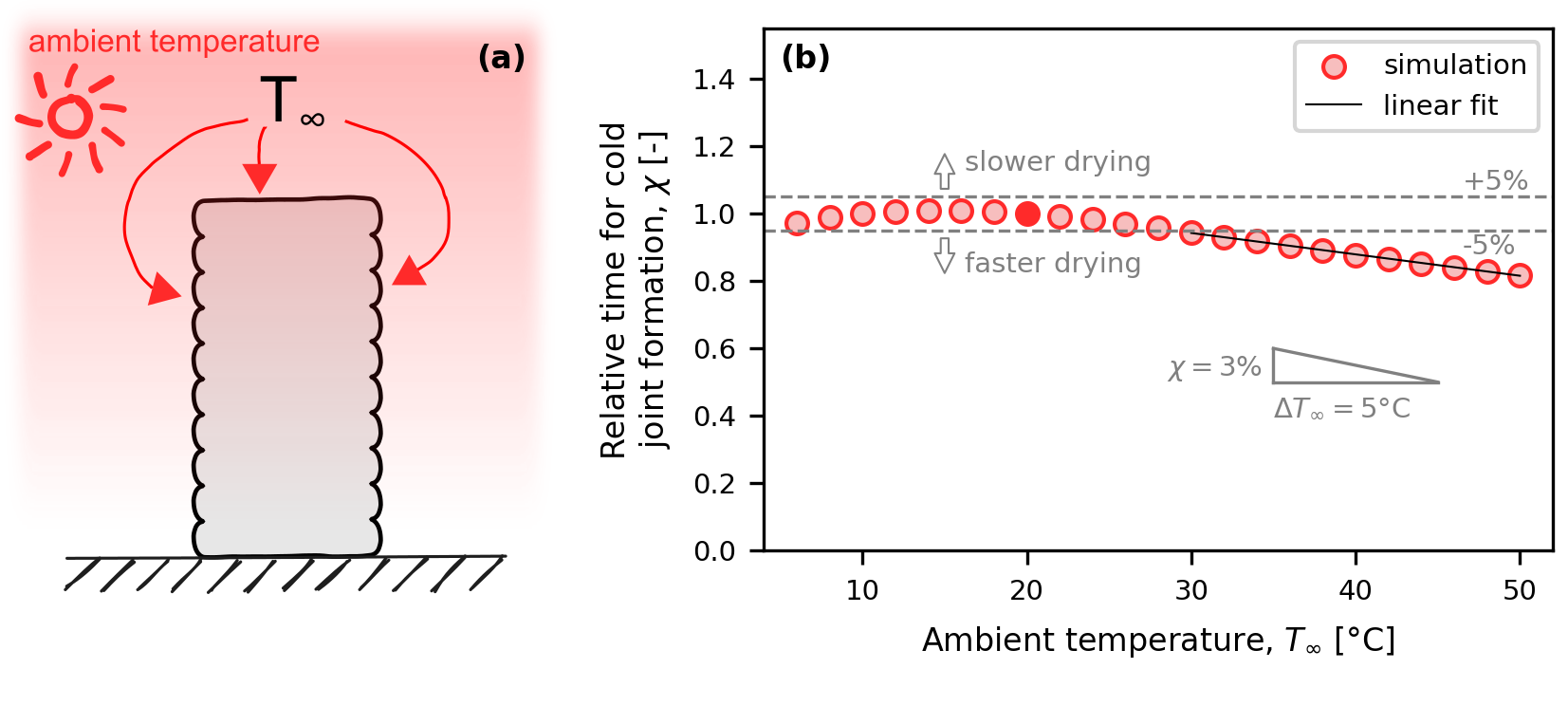}
	\caption{Influence of the ambient temperature on the concrete surface drying rate. (a) Schematic illustration of the affecting temperature. (b) Computed relative time to cold joint formation $\chi$ for a variation in ambient temperature.}
	\label{fig:factors_ambient_temp}
\end{figure}

\begin{figure}[htb!]
    \centering
    \includegraphics[scale=1]{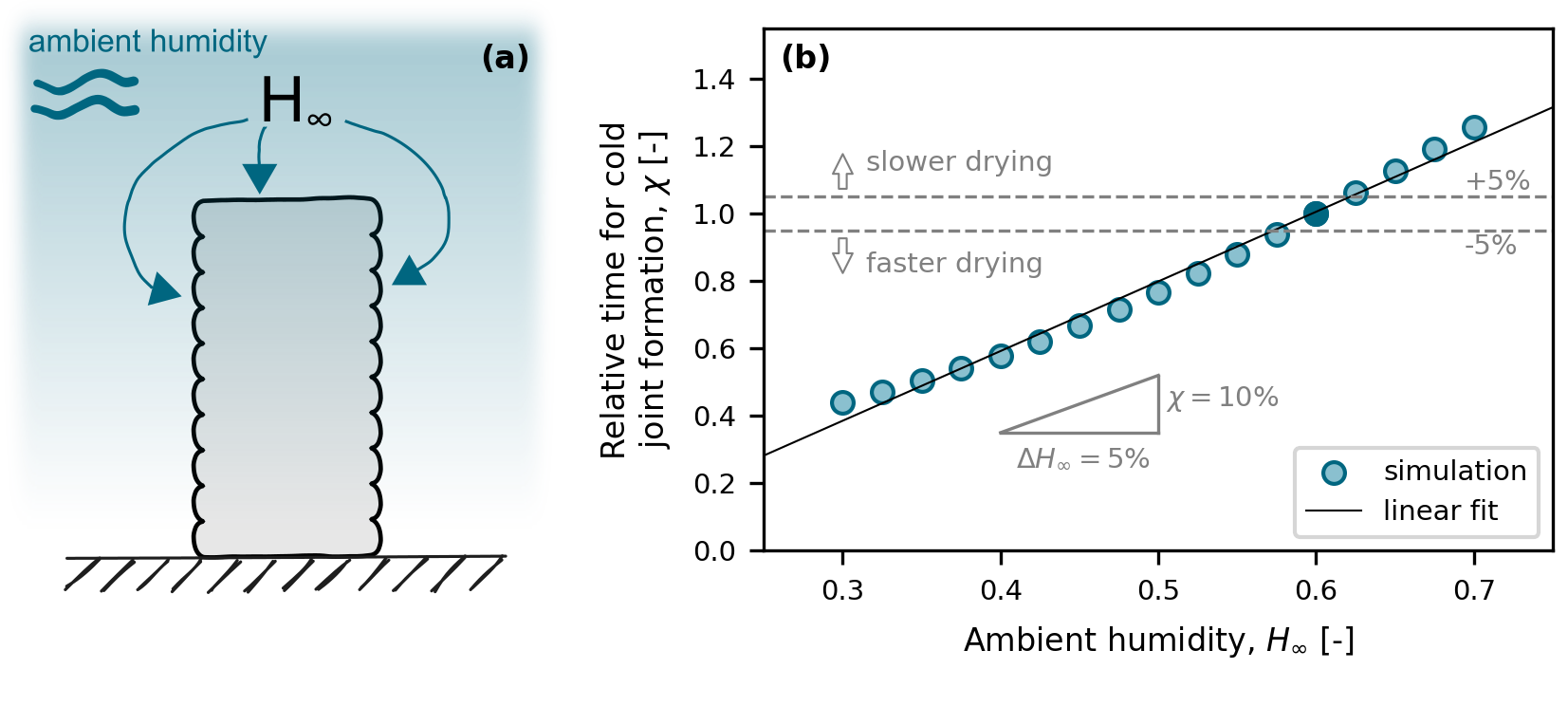}
	\caption{Influence of the ambient humidity on the concrete surface drying rate. (a) Schematic illustration of the affecting humidity. (b) Computed relative time to cold joint formation $\chi$ for a variation in ambient humidity.}
	\label{fig:factors_ambient_humi}
\end{figure}

The influence of the surrounding environment on the concrete structure comprises the contributions of ambient temperature and humidity, both of which fluctuate jointly throughout the day and season.
While their effects on the structure are coupled in real-life scenarios, the computational model enables us to dissect their impact and consequences separately, see~\reffig{factors_ambient_temp} and~\ref{fig:factors_ambient_humi}.
\FloatBarrier
\FloatBarrier

We consider an ambient temperature fluctuation in the range $T_\infty \in [6, 50]^{\circ}$C, corresponding to globally foreseeable extreme temperatures encountered during 3D concrete printing.
Calculated results for $\chi$ indicate two drying regimes, see~\reffig{factors_ambient_temp}(b).
For typical temperatures in the range between 6 to 30$^{\circ}$C, we observe no influence on the surface drying rate within $\pm5\%$ of $t_{cj}$.
Such temperatures are commonly present if the structure is shielded from direct sun exposure, as is usual in production halls and laboratories.
For ambient temperatures above $30^{\circ}$C, we observe $\chi<1$, denoting an accelerated cold joint formation rate.
In particular, $\chi$ decreases by $\approx 3\%$ per $5^{\circ}$C.
In summary, calculated results demonstrate that isolated temperature fluctuation within common temperature limits does not impact the kinetics of cold joint formation.

Ambient humidity has a strong effect on surface drying kinetics.
The considered humidity range $H_{\infty}\in [0.3, 0.7]~[-]$ covers the common limits encountered throughout the year.
Calculated results shown in~\reffig{factors_ambient_humi}(C) indicate a linear dependence between $H_{\infty}$ and $\chi$, namely $\chi \approx 10\%$ per 5\% $H_{\infty}$.
As a consequence, 3D printing in a dry ambient poses a significant challenge regarding accelerated surface drying, effectively shortening the time to cold joint formation.

Summarizing the effect of the ambient environment, low ambient humidity and high ambient temperature accelerate surface drying and increase the risk of cold joint formation.
While shielding the structure from the sun helps to lower the environmental temperature, active control of ambient humidity on the building site or laboratory is probably just reserved for niche applications.

\subsection{Humidity and heat exchange with environment}\label{sec:humExchange}
\begin{figure}[htb!]
    \centering
    \includegraphics[scale=0.85]{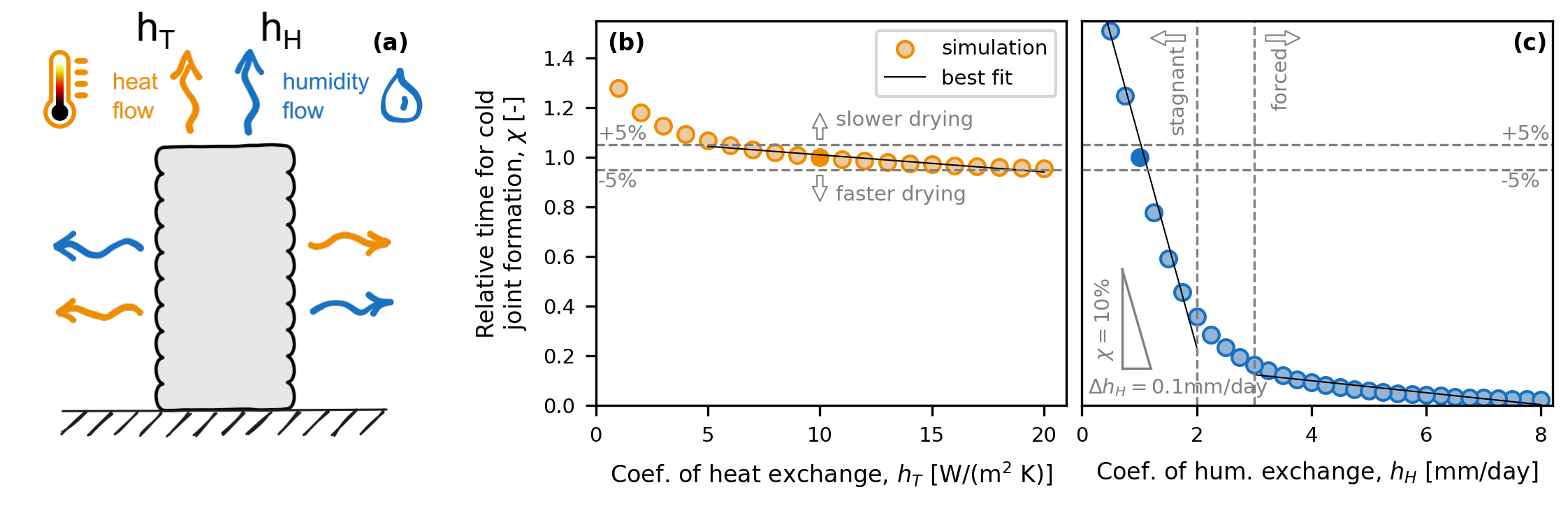}
	\caption{Influence of the heat and humidity exchange between the structure and its environment on the concrete surface drying rate. (a) Schematic separation of both phenomena. (B \& C) Computed relative time to cold joint formation for a variation of heat and humidity transfer coefficients, respectively.}
	\label{fig:factors_interaction}
\end{figure}

The uncovered surface of extruded concrete structures facilitates the transfer of heat and humidity from the structure to the environment, see illustration in~\reffig{factors_interaction}.  
Similar to~\refsec{ambient_conditions}, we here separately investigate the contribution of each phenomenon to the surface drying rate.
\FloatBarrier

We characterize the temperature exchange between the structure and its environment using the heat transfer coefficient $h_T$, see~\reffig{factors_interaction}(a).
For exposed sides, $h_T \in [1,20]$~W/(m$^2$~K), which represent typical values used for unshielded concrete freely exposed to the environment~\cite{Faria:2006, Smilauer:09, Jendele:2014, daSilva:2015}.
Calculated results of $\chi$ shown in~\reffig{factors_interaction}(b) indicate no noticeable influence on the surface drying rate for $h_T > 5$~W/(m$^2$~K), implying that heat exchange from an uninsulated concrete surface does not accelerate cold joint formation.
A time prolongation in their formation can be effectively achieved by a reduction of $h_T$ below 5~W/(m$^2$~K), indicating that the addition of a heat-insulating layer atop the extruded filament surface would help slow down surface drying.
Assuming $h_T=10$~W/(m$^2$~K) to represent the heat exchange between the naked concrete structure and the ambient environment, the addition of \textit{at least} 4 mm of polystyrene ($\lambda=0.033$~W/(m~K)) would be sufficient to delay their formation.

For humidity exchange, we consider a transfer coefficient $h_H \in [0.25, 8.00]$~mm/day.
The lower end of the $h_H$ range corresponds to gradual drying under stagnant conditions, e.g.\ structures positioned in a protected environment.
The upper range of the $h_H$ interval describes a surface exposed to a moving airflow.
Results of drying simulations shown in~\reffig{factors_interaction}(C) indicate a bilinear drying regime with a transition around $h_H = 2-3$~mm/day.
Under stagnant environmental conditions ($h_H < 2$~mm/day), the rate of surface drying is strongly affected by the chosen $h_H$ with $\chi \approx 10\%$ per 0.1~mm/day.
For extrusion with structures exposed to forced airflow, rapid drying rate becomes a major threat.
In such conditions, a stable $\chi \approx 0.05-0.15$ denotes a very fast cold joint formation.
We therefore emphasize that precise quantification of $h_H$ is crucial for accurate modeling surface drying rate.

As a practical recommendation, our modeling results indicate that any protective measure applied to the exposed concrete surface strongly affects its drying rate. 
Sealing the open surface with an impenetrable foil or spraying an evaporation retardant will efficiently increase $\chi$, prolonging the time before a cold joint forms.
3D printing under ambient conditions with active air circulation should be avoided or, if imperative, the extrusion should be carried out rapidly, minimizing the time between deposition of individual layers.

\subsection{Material state}\label{sec:material_state}
\begin{figure}[htb!]
    \centering
    \includegraphics[scale=0.85]{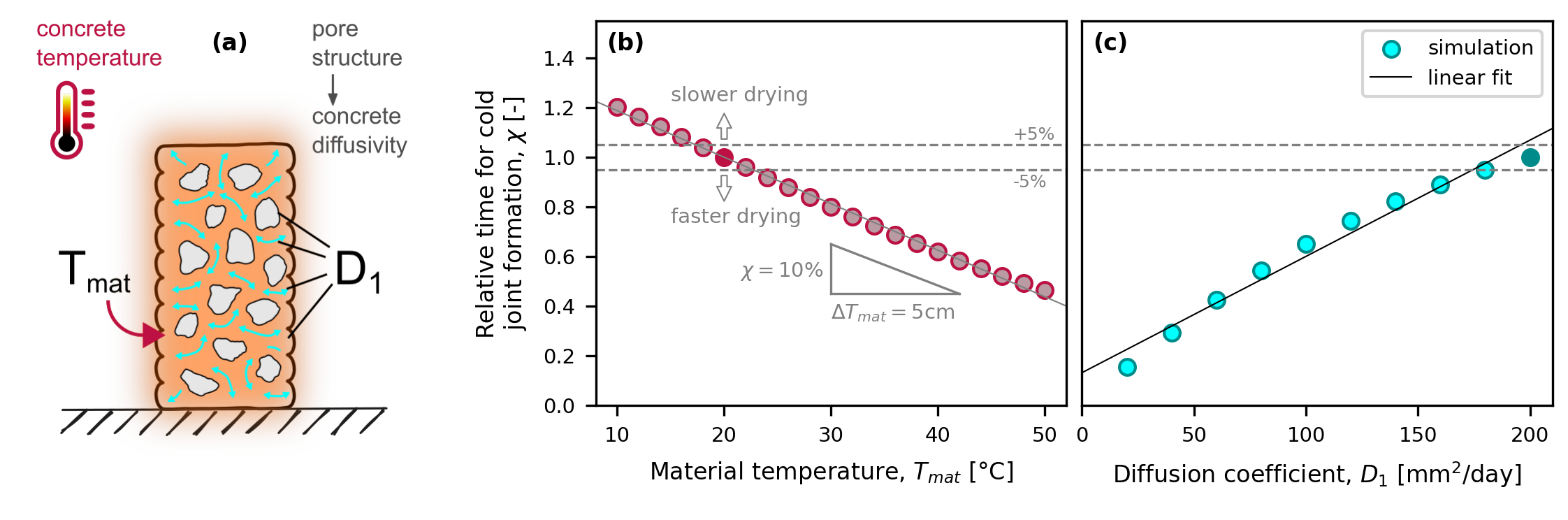}
	\caption{Influence of a material state described by the material temperature and diffusivity on the rate of cold joint formation. (a) Schematic separation of both phenomena. (B \& C) Computed relative time to cold joint formation for a variation in material temperature and diffusivity, respectively.}
	\label{fig:factors_material}
\end{figure}

Hydrating concrete generates measurable heat from the reacting binder, which is concurrently accompanied by the development of a pore network on the microstructural scale.
Here we investigate how the material temperature and the humidity diffusivity through the porous microstructure influence the surface drying rate, see~\reffig{factors_material}(a).
\FloatBarrier

In our fictitious concrete structure, the initial material temperature $T_{mat}$ corresponds to a temperature resulting from raw material mixing, pumping, extrusion, and partial cement hydration.
This processing history results in an increase of $T_{mat}$, offsetting it from $T_{\infty}$ at the model evaluation time.
We vary $T_{mat} \in [10, 50]^{\circ}$C in our drying simulations while maintaining $T_\infty = 20^{\circ}$C.
Calculated results indicate a linear relationship between $T_{mat}$ and $\chi$, see~\reffig{factors_material}(b), with $\chi \approx 10\%$ for every 5\degC in $T_{mat}$.
Notably, it is interesting to point out that if the concrete temperature is cooler than the ambient environment, this effectively prolongs the time necessary for cold joint formation.
This can be conveniently achieved by cooling down water with ice before mixing, rendering this recommendation easily applicable to every building site or laboratory.

The exposed concrete surface can be re-saturated by water diffusion through the internal pore network.
To account for this phenomenon, we use the coefficient $D_1$, which describes the rate of humidity diffusion through the complex pore structure.
We note that $D_1$ depends on the cement fineness distribution and w/c ratio used~\cite{Hlobil:2022cementPSD, Hlobil:2023DiB} and is affected by the ongoing microstructural development, but consider a constant $D_1$ value for simplicity in the simulations.

For the model evaluation, we assume a value $D_1 \in [20,200]$~mm/day.
Low values of $D_1$ indicate a dense microstructure with unconnected fine pores, hindering humidity transport from the structural core to the exposed outer surface.
This characterizes a microstructure of mature concrete.
The opposite spectrum of $D_1$ values corresponds to a highly porous medium with interconnected pores, i.e.\ a fresh concrete state, whose microstructure facilitates humidity transport within the material.
Simulated results indicate an almost linear relationship between $D_1$ and $\chi$.
Low values of $D_1$ result in a rapid formation of a dry, thin surface layer as the material cannot transport the internal humidity, shortening the time to cold joint formation.
High values of $D_1$ denote a highly permeable material that allows for a steady redistribution of internal water to the drying surface, effectively prolonging the time to cold joint formation.
However, providing recommendations for selecting a unique value of $D_1$ for a particular generic extruded structure is beyond the scope of this manuscript.

In summary, the chosen value for material diffusivity $D_1$ has a pronounced effect on the rate of surface drying.
However, since many factors influence the diffusivity through the pore network, it becomes impossible to set a recommendable initial value for model predictions.
The chosen value of $D_1$ serves only for qualitative assessment of the cold joint formation but can be adapted if a specific value becomes obtainable by other means.

\subsection{Structural geometry}
\begin{figure}[htb!]
    \centering
	\includegraphics[scale=0.85]{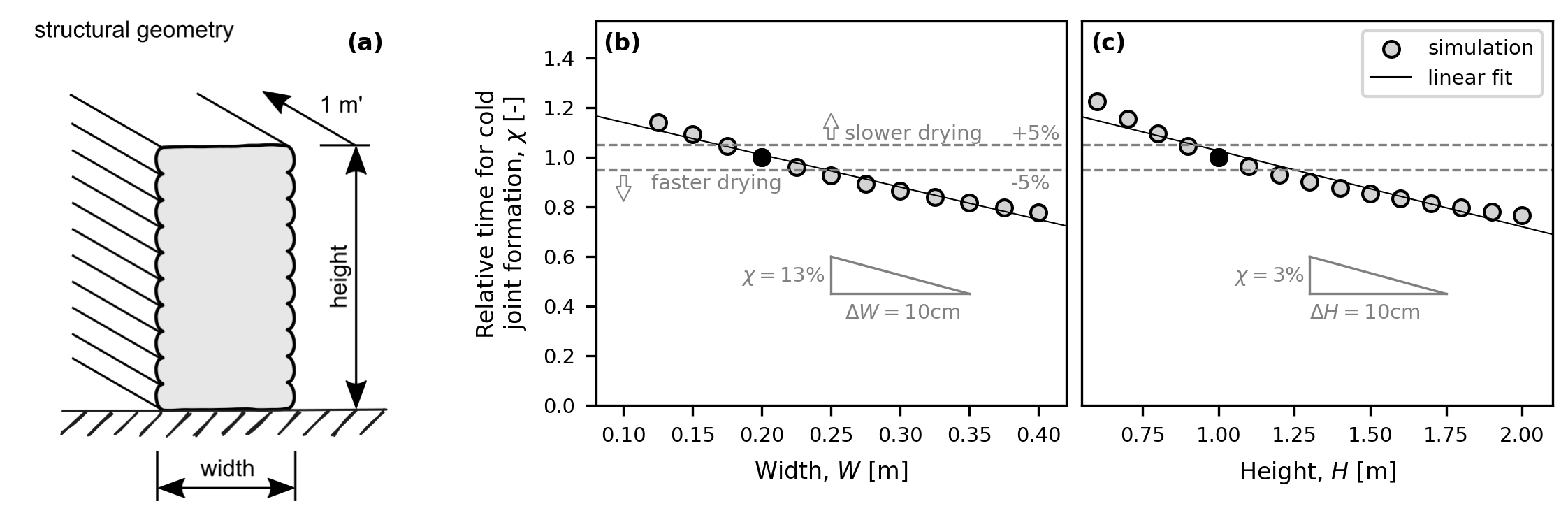}
	\caption{Influence of the structural geometry on the rate of cold joint formation.}
	\label{fig:factors_geometry}
\end{figure}

Extruded concrete structures typically exhibit a wall-like geometry with a very large height-to-thickness ratio.
While the extrusion nozzle typically determines the cross-sectional width (i.e., wall thickness), the structural height gradually increases following the sequential deposition of individual filaments.
We present the results of simulated drying kinetics for variable structural geometry below, see~\reffig{factors_geometry}.
\FloatBarrier

In our numerical study on the fictitious concrete structure, we varied both structural dimensions over expectable ranges achievable in the construction practice, i.e.\ $W \in [0.1,0.4]$~m and $H \in [0.7,2.0]$~m.
Calculated results show an approximately linear dependency of both structural dimensions with $\chi$, in particular, $\chi \approx 3\%$ for $\Delta H = 0.1$~m and $\chi \approx 13\%$ for $\Delta W=0.1$~m.
Based on purely geometric considerations, an increase in any structural dimension proportionally increases the exposed surface area and cement content within the bulk concrete; see discussion in~\refsecs{humExchange}{material_state}.
Both these factors shorten the time to form a cold joint.

In summary, the effect of structural geometry on the drying rate is non-negligible.
However, we acknowledge that the geometry of extruded structures is typically constrained by civil engineering and architectural requirements or technological limitations.
The presented model, however, allows a rapid assessment of any arbitrary structure with a wall-like geometry, making it a suitable tool for assessing the threat of cold joint formation for a given design.

\section{Conclusions}
Cold joints in extruded concrete structures form once the exposed surface of a deposited filament dries prematurely and gets sequentially covered by a layer of fresh concrete.
This creates a material heterogeneity which lowers the structural durability and shortens the designed service life.

This paper introduces a computational model for assessing the risk of cold joint formation that accounts for the extruded structural geometry, structural interaction with the environment, and ambient conditions.
The model provides, as an output, the time for cold joint formation, which allows us to assess the severity of surface drying for a given combination of input parameters.

The main conclusions from our study can be summarized as follows:
\begin{itemize}
    \item \textit{Ambient conditions:} Low ambient humidity $H_{\infty}$ and high ambient temperature $T_{\infty}$ shorten the time to cold joint formation at a rate of $10\%$ for $\Delta H_{\infty}= 5\%$ and $3\%$ for $\Delta T_{\infty}=5^{\circ}$C if $T_{\infty}>30^{\circ}$C, respectively.
    While shielding the structure from direct sun exposure helps to control the ambient temperature, a regulated increase of ambient humidity on the building site or laboratory is probably reserved for niche applications.
    \item \textit{Heat exchange with environment:} Isolated heat exchange between a naked concrete structure and its environment does not accelerate the surface drying rate. 
    Adding a thermal insulation layer (e.g., \textit{at least} 4 mm of polystyrene) atop the exposed surface effectively delays the time for cold joint formation.
    \item \textit{Humidity exchange with environment:} Surface drying is strongly affected by humidity exchange between the structure and the environment.
    The humidity coefficient $h_H$ increases the drying rate by 10\% for $\Delta h_H=0.01$~mm/day in a stagnant environment.
    Sealing the open concrete surface with water vapor barriers, temporarily adding impenetrable foils, or regular spraying with water provides an effective means for keeping the concrete surface from prematurely drying.
    3D printing outdoors (under ``open air” conditions) should be avoided or, if imperative, carried out at least uninterrupted, minimizing the time between the extrusion of individual layers.
    \item \textit{Material temperature:} Concrete temperature $T_{mat}$ strongly affects the surface drying and shortens the time to cold joint formation at a rate of $10\%$ for $\Delta T_{mat} =5^{\circ}$C.
    Cooling the mixing water before extrusion provides a very efficient and effective means of prolonging the time for cold joint formation.
    \item \textit{Material pore structure:} The material diffusivity remains an outcome of the microstructural development rather than an input parameter for a concrete mix design. 
    Concretes with a denser microstructure exhibit a lower diffusivity, translating into a faster cold joint formation due to the constrained water supply from the structural interior onto the drying surface.
    \item \textit{Structural geometry:} Structural geometry affects the drying rate as $3\%$ for $\Delta H = 0.1$~m and $13\%$ for $\Delta W=0.1$~m.
    The presented model allows the user to verify and compare the time for cold joint formation for any wall-like structural geometry and provides a relative time to cold joint formation.
\end{itemize}

\section{CRediT authorship contribution statement}
\textbf{Michal Hlobil:} Conceptualization, Methodology, Formal analysis, Investigation, Data curation, Writing - Original Draft, Visualization;
\textbf{Luca Michel:} Data curation, Formal analysis, Writing - Review \& Editing;
\textbf{Mohit Pundir:} Software, Writing - Review \& Editing;
\textbf{David S.\ Kammer:} Writing - Review \& Editing, Resources, Supervision, Project administration, Funding acquisition

\section{Declaration of Competing Interests}
The authors declare that they have no known competing financial interests or personal relationships that could have appeared to influence the work reported in this paper.

\section{Data availability}
The code used for the numerical simulations is available on \href{https://gitlab.ethz.ch/cmbm-public/papers-supp-info/2024/thermo-hygro-model-cold-joints}{ETHGitlab} and the generated simulation data has been deposited in the ETH Research Collection database under accession code \href{https://doi.org/10.3929/ethz-b-000675875}{10.3929/ethz-b-000675875}.

\section{Acknowledgement}
D.\ S.\ K., M.\ H., and L.\ M.\ acknowledge support by the Swiss National Science Foundation under grant 200021\_200343.
We are thankful to Yaqi Zhao and Dr.~Antoine Sanner for the helpful discussion and to Mikey Ronen for his feedback on the manuscript.

\appendix
\section{Derivation of governing equations}\label{sec:DerivationDiffEq}

\subsection{Transient heat conduction in concrete}\label{sec:transient_heat_conduction}
The temperature field within a hydrating concrete structure is highly non-uniform.
While cement hydration internally generates heat throughout the structure, the exposed surface is cooled down by heat exchange with the ambient environment.
This phenomenon may be conveniently modeled by formulating a transient heat conduction problem with an internal heat source over a given domain.
The governing equation for the thermodynamic equilibrium defined for a differential element of the structure requires
\begin{linenomath}
	\begin{eqnarray}
		\nabla \cdot \overline{\boldsymbol{q}}_T + \frac{\partial Q}{\partial t} = 0,
		\label{eq:heat_conduction}
	\end{eqnarray}
\end{linenomath}
where $\overline{\boldsymbol{q}}_T$ is the heat flux emerging from heat conduction in [J/(s~m$^2)]$ and $Q$ is the heat accumulated within a unit volume of concrete in [J/m$^3$].

For statistically homogeneous and isotropic materials, such as plain concrete investigated in the present study, the heat flux vector $\overline{\boldsymbol{q}}_T$ can be expressed using Fourier's law of heat conduction as a negative vector gradient of temperature, namely as 
\begin{linenomath}
	\begin{eqnarray}
		\overline{\boldsymbol{q}}_T = - \lambda_{mat}(\alpha) \, \nabla T,
		\label{eq:Fourier's_law}
	\end{eqnarray}
\end{linenomath}
where $\lambda_{mat}(\alpha)$ represents the heat conductivity of concrete in [J/(s~m~K)], reflecting the microstructural development described by $\alpha$.

The total heat $Q$ accumulated within a unit volume comprises two heat rate contributions: accumulated from conduction $Q_c$ and released from hydration $Q_h$, both given in [J/m$^3$], given as
\begin{linenomath}
	\begin{eqnarray}
		\frac{\partial Q}{\partial t} = \frac{\partial Q_c}{\partial t} - \frac{\partial Q_h}{\partial t} = \frac{\partial Q_c}{\partial T} \, \frac{\partial T}{\partial t} - \frac{\partial Q_h}{\partial t} = c_v \, \frac{\partial T}{\partial t} - \frac{\partial Q_h}{\partial t},
		\label{eq:Qsplit}
	\end{eqnarray}
\end{linenomath}
where $c_v$ stands for the volumetric heat capacity of concrete in [J/(m$^3$~K)].
This follows from multiplying the specific heat capacity of the material $c_{mat}$ in [J/(kg~K)] by its volumetric mass density $\rho_{mat}$ in [kg/m$^3$].

\subsection{Transient humidity conduction in concrete}\label{sec:derivHumidity}
Free (chemically unbound) water in the capillary pore network plays a central role in microstructural formation.
Here, we consider the relative humidity $H_{mat}$ as an equilibrated humidity of the gaseous phase and the interstitial liquid phase present within the pore network of the cementitious microstructure.
After the initial stages of cement hydration, the ongoing microstructural development becomes controlled by water availability from mutually unconnected pores, with water diffusion serving as a driving force for the supply of ionic species.
Thus, we formulate the conduction of humidity as a transient diffusion problem with $H_{mat}$ acting as the driving potential for the moisture field and with negligible moisture convection through pores, given as 
\begin{linenomath}
	\begin{eqnarray}
		\nabla \cdot \overline{\boldsymbol{q}}_H + \frac{\partial H_{mat}}{\partial t} =   - \frac{\partial H_s}{\partial t} +  \kappa \, \frac{\partial T_{mat}}{\partial t},
		\label{eq:diffusion_transport}
	\end{eqnarray}
\end{linenomath}
where $\overline{\boldsymbol{q}}_H$ represents the humidity flux in~[m$^2$/s], the sink term $H_s$ accounts for self-desiccation of the material, and $\kappa = \frac{\partial H_{mat}}{\partial T_{mat}}$ is the hygrothermic coefficient.
Ba\v{z}ant and Najjar~\cite{Bazant:1972MaS} propose
\begin{linenomath}
	\begin{eqnarray}
		\kappa = 0.0135 \, H_{mat} \frac{1-H_{mat}}{1.25-H_{mat}} \,.
		\label{eq:kappa}
	\end{eqnarray}
\end{linenomath}
The self-desiccation follows as a drop in moisture content $w_h$ due to ongoing cement hydration, given by
\begin{linenomath}
	\begin{eqnarray}
		w_h = w_{c,pot} \, m_c \, \alpha,
		\label{eq:w_h1}
	\end{eqnarray}
\end{linenomath}
where $w_{c,pot}$ stands for the amount of chemically-bound water necessary for complete cement hydration, expressed as $w_{c,pot}= 0.23$~kg of water per kg of cement used~\cite{Neville:1997}, and $m_c$ stands for the mass of cement used in 1~m$^3$ of concrete.
The corresponding decrease of relative humidity due to self-desiccation $H_s$ amounts 
\begin{linenomath}
	\begin{eqnarray}
		H_s = \frac{w_h}{k} ,
		\label{eq:w_h2}
	\end{eqnarray}
\end{linenomath}
where $k=\frac{\partial w}{\partial H_{mat}}$ stands for the moisture capacity of the material in [kg/m$^3$] determined as the slope of the desorption isotherm, assumed here as $320$~kg/m$^3$.

The constitutive law defines the humidity flux as a negative vector gradient of relative humidity $H_{mat}$ given as
\begin{linenomath}
	\begin{eqnarray}
		\overline{\boldsymbol{q}}_H = - D_H \nabla H_{mat},
		\label{eq:moisture_flux}
	\end{eqnarray}
\end{linenomath}
where $D_H$ represents the diffusivity coefficient in [m$^2$/s].
Several formulations for $D_H$ have been proposed, as summarized by the \textit{fib} Model Code 2010~\cite{fibMC2010}.
Here, we consider the formulation proposed by Ba\v{z}ant and Najjar~\cite{Bazant:1972MaS} in the form of
\begin{linenomath}
	\begin{eqnarray}
		D_H = D_1 \, \left( \gamma_0 + \frac{1-\gamma_0}{1+ {\left(\frac{1-H_{mat}}{1-H_c}\right)}^n } \right),
		\label{eq:diffusivity}
	\end{eqnarray}
\end{linenomath}
with parameters taken from~\cite{Sovjak:2018} as $D_1=200$~mm$^2$/day as the value of diffusivity at full pore saturation ($H_{mat}\to 1$), $\gamma_0 = \frac{D_0}{D_1}=0.05$~[--] with $D_0$ as the minimum $D_H$ for $H_{mat}\to 0$, $H_c=0.75$ as the relative pore humidity for $D_H = 0.5 \, D_1$, and $n = 12$ characterizing the drop in spread of $D_H$.
The value of $D_1$ can be estimated for mature concrete as a function of its average strength $f_{cm}$~\cite{fibMC2010} as 
\begin{linenomath}
	\begin{eqnarray}
		D_1 = \frac{10^{-8}}{f_{cm}-8},
		\label{eq:D1-fc}
	\end{eqnarray}
\end{linenomath}
which provides an effective range of $D_1$ between $[200, 20]$~mm$^2$/day for typical concrete strengths $f_{cm}$ between $[12, 50]$~MPa.
The impact of $D_1$ on simulation results is shown in detail in~\refsec{cj_formation}.

\section{Parameters for the affinity-based hydration model}\label{appendix:Calibration_affinity-model}
The analytical expression for the affinity-based hydration model given in~\refeq{affinity} uses $\alpha_{\infty}$ as the maximum-reachable degree of hydration in cement paste. 
This parameter depends on the initial water-to-cement ratio $w/c$ and the curing regime considered~\cite{Brouwers:2004}.
As such, $\alpha_{\infty}$ takes the form of
\begin{linenomath}
	\begin{eqnarray}
		\alpha_{\infty} = 
		\begin{cases}
			\frac{w/c}{0.42} \leq 1& \quad \text{(sealed)}, \\
			\frac{w/c}{0.36} \leq 1 & \quad \text{(saturated)}.
		\end{cases}
		\label{eq:DoHmax}
	\end{eqnarray}
\end{linenomath}
However, the practically reachable limit for $\alpha_{\infty}$ for ordinary Portland cements under common curing conditions and in reasonable time-scales (up to years of hydration under normal temperature) amounts to $\approx 0.85$~[-], which will be used for model evaluation.

\reftab{kinetics} summarizes the parameters of the affinity-based hydration model that best fit the experimental data from isothermal calorimetry, as shown in~\reffig{affinity-model-predictions}.
\begin{table}[htp!]
\centering
	\caption{Calibrated parameters for affinity-based hydration model for isothermal calorimetry measurements of various cement pastes.}
	\label{tab:kinetics}
\begin{tabular}{|l|c|c|c|c|c|}
	\hline
	\hline
	& CEM~I Ladce  & CEM~I Ladce   & CEM~II B/LL  & CEM~I Ladce  \\
 	& (391~m$^2$/kg) & (391~m$^2$/kg) + 20\% quartz   & 32.5 R Hranice & (273~m$^2$/kg)  \\
	\hline
	$B_1~[s^{-1}]$ & $142 \cdot 10^{-6}$ & $109 \cdot 10^{-6}$ & $67 \cdot 10^{-6}$ & $104 \cdot 10^{-6}$ \\
	$B_2~[-]$  & $408\cdot 10^{-5}$ & $1210\cdot 10^{-5}$ & $1829\cdot 10^{-5}$ & $477\cdot 10^{-5}$ \\
	$\eta~[-]$ & $610\cdot 10^{-2}$ & $550\cdot 10^{-2}$ & $332\cdot 10^{-2}$ & $608\cdot 10^{-2}$ \\
	$\alpha_{\infty}~[-]$ & 0.85 & 0.85 & 0.85 & 0.85 \\
	$Q_{tot}~[J/g]$ & 532 & 426 & 330 & 512 \\
	\hline
	\hline
\end{tabular}
\end{table}

\section{Parameters for computational model evaluation}\label{appendix:Params4modelEvaluation}
\reftab{params4sims} summarizes the parameters used for the analytical hydration model, which are used for model verification in Sections~\ref{sec:verif_temperature}, \ref{sec:verif_drying}, and \ref{sec:cj_prediction}.
\begin{table}[htp!]
\centering
	\caption{Used parameters for affinity-based hydration model for simulations.}
	\label{tab:params4sims}
\begin{tabular}{|l|c|c|c|c|}
	\hline
	\hline
	& Section~\ref{sec:verif_temperature} & Section~\ref{sec:verif_drying}   & Section~\ref{sec:cj_prediction}  \\
	\hline
	$B_1~[s^{-1}]$  & $134 \cdot 10^{-6}$ & $142 \cdot 10^{-6}$ & $142 \cdot 10^{-6}$ \\
	$B_2~[-]$  & $689 \cdot 10^{-5}$ & $408 \cdot 10^{-5}$ & $408 \cdot 10^{-5}$  \\
	$\eta~[-]$ & $573 \cdot 10^{-2}$ & $610 \cdot 10^{-2}$ & $610 \cdot 10^{-2}$  \\
	$\alpha_{\infty}~[-]$  & 0.85 & 0.85  & 0.83  \\
	$Q_{tot}~[J/g]$ & 520 & 532 & 532 \\
	\hline
	\hline
\end{tabular}
\end{table}

\bibliography{myLiterature.bib}

\end{document}